\documentclass[preprint,showpacs,preprintnumbers,amsmath,amssymb,nofootinbib]{revtex4}

\usepackage{graphicx}
\usepackage{dcolumn}
\usepackage{bm}

\begin{document}
\title{\bf On the laser frequency stabilization by locking to a LISA arm}

\author{Massimo Tinto}
\email{Massimo.Tinto@jpl.nasa.gov}
\altaffiliation [Also at: ]{Space Radiation Laboratory, California
  Institute of Technology, Pasadena, CA 91125}
\affiliation{Jet Propulsion Laboratory, California Institute of
Technology, Pasadena, CA 91109} 

\author{Malik Rakhmanov}
\email{malik@phys.ufl.edu}
\affiliation{Department of Physics, University of Florida,
  Gainesville, FL 32611}

\date{\today} 
\begin{abstract}
  LISA is an array of three spacecraft flying in an approximately
  equilateral triangle configuration, which will be used as a
  low-frequency detector of gravitational waves. Recently a technique
  has been proposed for suppressing the phase noise of the onboard
  lasers by locking them to the LISA arms. In this paper we show that
  the delay-induced effects substantially modify the performance of
  this technique, making it different from the conventional locking 
  of lasers to optical resonators. We analyze these delay-induced 
  effects in both transient and steady-state regimes and discuss their 
  implications for the implementation of this technique on LISA.
\end{abstract}

\maketitle

The Laser Interferometer Space Antenna (LISA) \cite{1} is a
space-borne interferometer that will use coherent laser beams
exchanged between three widely separated spacecraft to detect
low-frequency ($10^{-4}-1$ Hz) gravitational waves. Each spacecraft is
carrying two optical benches with lasers, optics, photo-detectors, and
drag-free proof masses. It has been shown that the time series of
Doppler shifts of the laser beams between spacecraft pairs, and those
between adjacent optical benches within each spacecraft, can be
combined, with suitable time delays, to cancel the otherwise
overwhelming laser phase noise.  This post-processing data technique
is known as time-delay interferometry (TDI) (see \cite{2} and
references therein).  Recently a method to reduce the laser phase
noise at the time of detection by using the arms of LISA has been
proposed by Sheard et al.  \cite{3}.  In this one-arm locking
technique a fraction of the beam from the laser on board of one
spacecraft interferes with a beam coherently retransmitted back by
another spacecraft, as shown in Fig. 1.  The phase difference of the
two beams forms an error signal for a control system feeding back to
the laser.  Originally proposed as an alternative to the TDI
technique, this method can be implemented in conjunction with TDI to
provide laser pre-stabilization, which will relax the requirements for
implementing the TDI technique.

The motivation for the one-arm stabilization method comes from the
conventional techniques of locking lasers to optical resonators, in
which the suppression of laser frequency noise is ultimately defined
by the stability of the resonator. In the case of LISA, this notion
becomes substantially modified by the delay-induced effects
originating from the time of flight within the LISA arm. Most
noteworthy of these effects is the prolonged ringdown discovered in 
numerical simulations \cite{3} and later confirmed in table-top
experiments \cite{Garcia04}. In this paper we provide a general theory 
for these delay-induced effects and analyze their impact on the 
performance of this stabilization technique.




Following \cite{3}, we neglect variations in the distance between the
spacecraft and thus assume that the duration of the photon round-trip
within the LISA arm ($\tau$) is constant: $\tau = 33.3$~s.  The
inverse of the photon round-trip within the LISA arm ($1/\tau)$ plays
a special role in this analysis and will be called free spectral range
(FSR) by analogy with optical resonators \cite{Verdeyen}. Our notations
are shown in Fig. 1: $p(t)$ is the free-running laser phase noise,
$q(t)$ is the residual phase noise, $r(t)$ is the error signal, and
$G(t)$ is the filtering function of the control system. The
time-domain equations for the closed-loop dynamics are
\begin{eqnarray}
   r(t) & = & q(t) - q(t - \tau),
              \label{def:r(t)} \\
   q(t) & = & p(t) - \int_0^t G(t-t') \; r(t') \, dt' ,
              \label{def:q(t)}
\end{eqnarray}
where we explicitly show that the feedback loop is closed at time
$t=0$. 

\begin{figure}[t]
   \centering\includegraphics[width=0.8\textwidth]{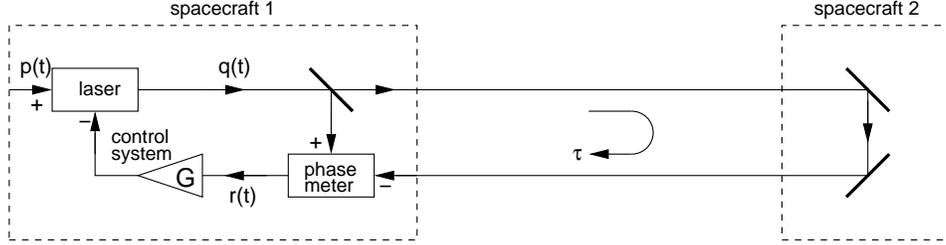}
   \caption{Schematic diagram of the one-arm laser stabilization 
     technique (adapted from \cite{3}).}
   \label{diagram}
\end{figure}

Consistency requires the laser to be running at least for the duration
of one photon round-time before the loop is closed, implying:
\begin{equation}
q(t) = p(t) \ \ \ , \ \ \ {\rm  for} \ t < 0 \ .
\label{boundary}
\end{equation}
By taking the Laplace transform of Eq.(\ref{def:r(t)}),
(\ref{def:q(t)}), and properly accounting for equation
(\ref{boundary}), we finally get
\begin{eqnarray}
   r(s) & = & q(s) - e^{-s \tau} \; q(s) - 
          \int_0^{\tau} e^{-st} \; p(t - \tau) \; dt ,\\
   q(s) & = & p(s) - G(s) \; r(s) .
\end{eqnarray}
Eliminating $r(s)$, we find that the residual noise in the
laser frequency stabilization loop is given by
\begin{equation}
   q(s) = C(s) \; p(s) + 
      C(s) \; G(s) \; \int_0^{\tau} e^{-st} \; p(t - \tau) \; dt ,
\label{closedLoop}
\end{equation}
where $C(s)$ is the closed-loop transfer function:
\begin{equation}
   C(s) = \frac{1}{1 + G(s)(1 - e^{-s \tau})} .
\end{equation}
Notice that the last term in Eq.(\ref{closedLoop}) is missing in the
analysis of \cite{3}. We now consider transient and steady-state
dynamics which are described by Eq.(\ref{closedLoop}). Henceforth we
assume that the feedback loop is stable and therefore all the
poles of $C(s)$ are on the left-half-plane of the Laplace domain.

Mismatch of the initial conditions (non-zero value of the error
signal) at the moment of closing the feedback loop causes start-up
transients \cite{3, Garcia04}, which can hinder the performance of the
one-arm locking. Here we provide simple formulas for the frequencies 
$f_n$ and relaxation times $\tau_n$ of these transients. They can
be found from the real and imaginary parts of the poles of the
closed-loop transfer function:
\begin{equation}
   s_n \equiv 2 \pi i f_n  - \frac{1}{\tau_n} .
\end{equation}
The closed loop transfer function $C(s)$ has poles
which are defined by the characteristic equation:
\begin{equation}
  1 - e^{-s \tau} = - \frac{1}{G(s)}.
\label{Charac}
\end{equation}
This equation has an infinite number of roots typical of dynamical 
systems with delay \cite{5}. For large gain, the roots can be found 
perturbatively in terms of inverse powers of $G$. Setting the right
hand side of Eq.(\ref{Charac}) to zero we obtain the zero-order 
approximation:
\begin{equation}
   \bar{s}_n = \frac{2\pi i}{\tau} \; n  ,
\end{equation}
where $n$ is integer. The next order approximation would be $s_n =
\bar{s}_n + \delta s_n$, where $\delta s_n$ is a small correction
vanishing in the limit of infinite gain. Simple algebra shows that
$\delta s_n \approx - [\tau G(\bar{s}_n)]^{-1}$.  Therefore, to first
order in $G^{-1}$, the poles of the closed-loop transfer function are
given by
\begin{equation}
   s_n = \frac{2 \pi i}{\tau} \; n - \frac{1}{\tau G(\bar{s}_n)}.
\end{equation}
Then the frequencies and relaxation times of the transients are
\begin{eqnarray}
   f_n & = & \frac{n}{\tau} - \frac{{\mathrm{Im}}\{
             G(\bar{s}_n)^{-1}\}}{2\pi\tau} ,\label{freqn} \\
   \tau_n & = & \frac{\tau}{{\mathrm{Re}} \{ 
                G(\bar{s}_n)^{-1} \} } . \label{taun}
\end{eqnarray}
Thus, we have shown that the transients form damped oscillations with 
frequencies fairly close to multiples of the free spectral range, and
their relaxation times are defined by the open-loop gain of the
control system. The amplitude of these transients can vary from lock
to lock, depending on the values of $p(0)$ and $p(-\tau)$. They can be 
significantly reduced by introducing a ramping function to the control
system which slowly increases the gain from zero to the designed value 
\cite{3,Garcia04}. 
However, the frequencies $f_n$ and relaxation times $\tau_n$ of the
transients are defined by the properties of the system and not the
initial conditions. To avoid prolonged ringdown of the transients, 
one can reduce $\tau_n$ by properly designing 
the open loop gain $G(s)$.

We now turn our attention to the steady-state dynamics of the closed
loop. Using the identity,
\begin{equation}
   C(s) \; G(s) = \frac{1 - C(s)}{1 - e^{-s \tau}} ,
\end{equation}
we can rewrite Eq.(\ref{closedLoop}) in the following equivalent
form:
\begin{equation}
   q(s) = C(s) \left[p(s) - P(s) \right] + P(s),
   \label{suppNoise}
\end{equation}
where we introduced a new quantity:
\begin{equation}
   P(s) = \frac{\int\limits_0^{\tau} e^{-st} \; 
      p(t - \tau) \; dt}{1 - e^{- s \tau}} .
\label{periodic}
\end{equation}
This is the Laplace transform of a periodic function of time \cite{4}, 
$P(t)$, which is determined by the repetition of the first $\tau$ seconds of the
free-running laser noise $p(t)$
\begin{equation}
   P(t + n \tau) = p(t) ,
\end{equation}
where $-\tau < t \leq 0$ and $n$ is integer.
Equation (\ref{suppNoise}) means that the free-running laser phase noise is 
compared with the periodic replica of its first $\tau$ seconds and the
difference between the two is suppressed by the loop gain $G(s)$. For
very large gain, $C(s)$ becomes negligible and
Eq.(\ref{suppNoise}) reduces to 
\begin{equation}
   q(s) = P(s) \ ,
   \label{limit}
\end{equation}
which describes the limiting performance of the one-arm locking
technique. In this case, the phase noise during the first $\tau$ 
seconds repeats itself indefinitely, thus becoming a periodic function
of time. The same conclusion can be drawn already from 
Eq.(\ref{def:r(t)}) which shows that the null point of the control 
system: $r(t) = 0$ implies periodicity of the residual noise: 
$q(t)=q(t - \tau)$.

The above calculation is based on the assumption of constant
round-trip-light-time which corresponds to perfect length reference. In this
situation, conventional stabilization of lasers to optical 
resonators would lead to an unlimited suppression of the laser 
phase noise. (In practice, the limitations would come from other 
noise sources such as the shot noise in the photo-detectors, which we
neglected in this analysis.)
As we have shown, the laser phase noise in the one-arm locking method
approaches a limit which is independent from the stability of the LISA
arm. Since the free-running laser phase noise is compared against
itself (at some earlier time) it can only be suppressed to the level
of its own stability within the first $\tau$ seconds.  This still
implies a suppression when viewed in the frequency domain.  Namely,
the power spectrum of the laser phase noise is no longer spread over a
large range of frequencies, rather it is concentrated within the
harmonics of the FSR. For finite gains, the peaks appear at
frequencies $f_n$ defined by the poles $s_n$ and are slightly offset
from the exact harmonics of the FSR. Furthermore, the peaks will have
a finite width and quality factors $Q_n = \pi f_n \tau_n$. Periodicity
of the residual noise suggests filtering methods which can done either
during detection (notch filtering) or in post-processing.
\vskip12pt\noindent
M.R. acknowledges support from the US National Science Foundation under 
grant PHY-0244902. This research was performed at the Jet Propulsion 
Laboratory, California Institute of Technology, under contract with 
the National Aeronautics and Space Administration.

\end{document}